# Domain-specific optimization and diverse evaluation of self-supervised models for histopathology

Jeremy Lai[1], Faruk Ahmed[1*], Supriya Vijay[1*], Tiam Jaroensri[1], Jessica Loo[2], Saurabh Vyawahare[2], Saloni Agarwal[1], Fayaz Jamil[1], Yossi Matias[1], Greg S. Corrado[1], Dale R. Webster[1], Jonathan Krause[1], Yun Liu[1], Po-Hsuan Cameron Chen[1], Ellery Wulczyn[1†], David F. Steiner[1†]

Affiliations:
[1]Google, Mountain View, CA
[2]Verily, South San Francisco, CA

*These authors contributed equally
†These authors co-supervised the work

## Abstract

Task-specific deep learning models in histopathology offer promising opportunities for improving diagnosis, clinical research, and precision medicine. However, development of such models is often limited by availability of high-quality data. Foundation models in histopathology that learn general representations across a wide range of tissue types, diagnoses, and magnifications offer the potential to reduce the data, compute, and technical expertise necessary to develop task-specific deep learning models with the required level of model performance. In this work, we describe the development and evaluation of foundation models for histopathology via self-supervised learning (SSL). We first establish a diverse set of benchmark tasks involving 17 unique tissue types and 12 unique cancer types and spanning different optimal magnifications and task types. Next, we use this benchmark to explore and evaluate histopathology-specific SSL methods followed by further evaluation on held out patch-level and weakly supervised tasks. We found that standard SSL methods thoughtfully applied to histopathology images are performant across our benchmark tasks and that domain-specific methodological improvements can further increase performance. Our findings reinforce the value of using domain-specific SSL methods in pathology, and establish a set of high quality foundation models to enable further research across diverse applications.

# Introduction

Computational pathology, and specifically the development of machine learning (ML) methods for the analysis of histopathology images has seen significant interest and progress in recent years[1–5] . However, image and data scarcity along with the large volume of expert annotations required to train task-specific ML models can limit the development, deployment, and generalizability of this promising technology. As ML for computer vision has advanced rapidly in recent years, there are emerging opportunities to explore the application of state-of-the-art approaches to computational pathology, potentially helping to address such issues of scalability and generalizability. One such category of techniques is self-supervised learning (SSL), which enables ML models to learn image features without the need for semantic labels acquired through human supervision or annotation[6–12]. SSL methods developed for natural images might be presumed to be applied readily to computational pathology. However, there are domain-specific aspects worth considering and histopathology involves visual features that are quite different from most natural images. Pathology images are typically in the form of whole slide images (WSIs) which are also significantly larger than most natural images and the semantic concepts in pathology span a large range in terms of the optimal resolution and field of view, from patch-level concepts (eg. specific regions and cellular patterns) to slide-level and case-level concepts (tissue orientation, microenvironment, and cellular and structural interaction).

A number of efforts have begun to explore the application of SSL to medical imaging[13] and histopathology, with promising results using both convolutional neural networks[14–17] and vision transformers (ViTs)[18–22]. Kang et al. report a preprint comparison of four SSL approaches and five benchmark tasks, finding that while no specific approach was necessarily best for all tasks, SSL models outperformed supervised pre-training on natural images[21]. Additional preprint reports also provide initial insights into ViT-based SSL approaches in pathology in regard to large models and a variety of downstream tasks[19,20,22]. Others have explored variations of the network architecture itself, including combining convolutional neural network and ViT architectures[23] or leveraging ViTs and SSL with hierarchical levels to learn robust representations of histopathologic images[24]. While these contributions and others have provided important insights, there are still many questions yet to be addressed in this rapidly evolving field such as the impact of training variations, data scale, and model size on learned representations and downstream applications.

To further advance the application of SSL to histopathology, we explore state-of-the-art SSL methods in combination with techniques from the computational pathology literature. This includes experiments with different SSL methods, data balancing, and pathology-specific augmentations such as image blur and cropping variations. Importantly, we utilize an extensive set of previously published benchmark pathology tasks and datasets to tune and evaluate across diverse tissue types and magnifications, thus enabling increased confidence in both the results and the generalizability of the resulting models. Our findings reinforce the value of using domain-specific SSL methods in pathology over pre-training using even the largest natural image pre-training sets available and provide further insights into utilization of SSL embeddings for patch-level, slide-level, and case-level tasks.

# Methods

## SSL Model Training

### Training Data

Training data consisted of hematoxylin and eosin stained (H&E) WSIs from The Cancer Genome Atlas (TCGA) accessed via https://portal.gdc.cancer.gov. TCGA is organized into individual studies representing different cancer types and images from 32 TCGA studies were included for training. Frozen slides from TCGA were excluded to maximize training on H&E images from formalin-fixed paraffin-embedded (FFPE) tissue, which are the most common intended target of downstream applications. TCGA data was also limited to images scanned at 20× as this is the highest magnification available for all TCGA studies. We did not use images scanned at 40× as this magnification is only available for a subset of the TCGA studies and introduces bias in terms of specific sites and scanners. Available WSIs were split into train (n=6249), tune (n=3079) and test splits (n=3065) by case to avoid any overlap of SSL training slides with slides used for downstream test sets containing TCGA data (noting that most patch-level evaluation test sets did not contain TCGA data, see SSL Model Evaluation section below).

For SSL model training, 60 million image patches (either 256×256 pixels or 512×512 pixels) were randomly sampled from the train split (with additional sampling variations as described in the "Data Sampling" of the Methods section below). To develop models robust to differences in magnification (ie. resolution or "zoom"), patches were sampled evenly across multiple magnifications, including 5× (~2 µm/pixel), 10× (~1 µm/pixel), and 20× (~0.5 µm/pixel).

### Modeling Overview

This work focused on utilization of a Vision Transformer (ViT) backbone (ViT-S and ViT-B)[7,25], with two different SSL methods: SimCLR[11,26] and MSN (Masked Siamese Networks)[10] as representative contrastive and non-contrastive SSL approaches, respectively. In preliminary experiments, we also experimented with a ResNet backbone and other methods such as masked autoencoders (MAE)[27] and DINO[6], but the ViT backbone with SimCLR and MSN performed best in initial linear probing evaluation and thus we used these approaches for our baseline SSL models ("SimCLR-base" and "MSN-base" respectively).

These baseline SSL models utilized the ViT-S/16 backbone, taking input image patches of 224 × 224 pixels (14 × 14 ViT patch tokens of 16 × 16 pixels each; output embedding vector of size 384 for each input image patch). Encoder backbones were initialized with weights from a ViT-S/16 trained on ImageNet-21k via the AugReg approach[25]. Final training details and hyperparameters for the baseline models are provided in Supplemental Table 1a.

Following establishment of the baseline SSL models, we explored a number of pathology-motivated methodological variations to further optimize SSL for histopathology. These included data augmentations, data sampling, training loss variations and the use of center embeddings. Details on these specific methods are described below. These variations were first added to the baseline SSL models individually. Variations that led to improvements were combined to produce our final SSL models. For combining individual variations, hyperparameters specific to each variation were re-tuned, and may differ from those selected for use of the individual variations alone.

## Data Augmentations

**Color augmentation** We utilized RandStainNA[21, 28] as it showed improved downstream performance in pilot experiments when used in conjunction with the typical color jittering operations of SSL data augmentation. Briefly, for every image, RandStainNA picks one of three color spaces — LAB, HSV, HED[29] — uniformly at random, and applies the Reinhard transformation[30] to fit the first and second-order color statistics of the image in the sampled color space to a target template. This target template is a sample from per-channel Gaussian fits to the color statistics from a random selection of training images in their respective color spaces. In preliminary experiments, we also explored perturbations in chrominance-space, and the GMM variant for RandStainNA proposed by Kang et al.[21] but consistently found RandStainNA to lead to best performance based on our linear probe metric.

**PathBlur: Poisson noise and JPEG artifacts** While application of blurring as a data augmentation strategy in pathology images can simulate out-of-focus patterns in pathology imaging (in addition to potentially encouraging the learning of higher-order structure or morphology), it also smooths away the natural artifacts of digital pathology imaging that arise from sensor noise and image compression[31]. In consideration of these natural artifacts, we explore an additional data augmentation method, referred to here as “PathBlur”, where we apply Poisson noise and JPEG artifacts (through an encode-decode step) after the application of Gaussian blur. Our hypothesis is that simulating such natural sources of noise during SSL might improve pathology-specific feature learning and enable additional robustness for downstream inference tasks. Poisson noise and JPEG artifact augmentations were also explored separately during exploration of individual and combined improvements for final models.

## Data Sampling

**Data rebalancing** Prior work suggests that balancing the data distribution is important for working with large datasets in the context of SSL[32]. This could be particularly important for large-scale pathology datasets, where the distribution across cell types and tissue morphology may be far from uniform.

To balance the original training dataset, we applied a cluster-based sampling approach. We trained an initial SSL model on the 60 million patch training dataset. Patch-level embeddings from this initial SSL model were used to define clusters using k-means clustering (using cosine distance as the similarity metric). All patches in the initial training dataset were assigned to the clusters. Then a new dataset was generated by sampling a fixed number of patches from each cluster for a total of 6 million patches.  A new SSL model was trained on this cluster-balanced dataset.

In initial experiments, we explored different numbers of clusters (between 1000 and 20000) and found k=18000 to result in best patch-level linear probe tune set performance. We also explored applying a second iteration of clustering and re-sampling based on the model from the first iteration, but did not observe any additional benefit. The best performing configuration was used for combining with the other training variations.

**Image crop overlap** Since semantic content in pathology image patches can potentially have a different spatial distribution than is found in natural image datasets, we explored modifications to the typical crop-and-resize data transformation in SSL approaches. Usually crops are randomly sampled from the image without consideration of overlap, since object-centric natural image datasets tend to exhibit a center-bias[33], which is likely to capture related semantic content in all crops.  We experimented with different degrees of overlap between crops used in the SSL training process. In particular, we ran a grid search where we explored a range of enforced overlap between crops (see Supplemental Table 1b) in the training configurations for both SimCLR and MSN approaches. The selected cropping strategy is depicted in Supplemental Figure 1.

## Training Loss Variations

**Hard negative mining (re-weighting loss)** We explore the use of hard negative mining in our SimCLR loss (NT-Xent) using a reweighting method described previously[34]. Briefly, the method works by using importance sampling on the set of negative images for an anchor image, up-weighting instances that are harder to separate from the anchor at any stage of training. We ramp up the strength of up-weighting across training steps, starting from an initial condition of no re-weighting.  We also experimented with debiasing the loss[35] to account for potential negative-set contamination that could impact learning, but in initial experiments we did not find debiasing to be beneficial for our linear probe tasks.

**SimCLR-MSN hybrid loss** In initial experiments, we noticed a trend of SimCLR performing better in tasks with more granular regions of interest (e.g. lung histologic subtyping and mitotic figure detection), whereas MSN performed better at the less granular tasks (e.g. TCGA study types and tertiary teaching hospital tissue types). This observation in addition to past studies examining the similarities of contrastive and non-contrastive SSL training[36,37] led us to experiment with the addition of a SimCLR loss (NT-Xent) to the MSN training. Specifically, each pair of teacher and student global views in the MSN training were treated as positive pairs in a batch with the rest of the batch's teacher and student global views serving as negative examples. We also tested using global and local views of the same image as positive pairs, but this did not perform as well as only applying the loss to global views.

### Center Embeddings

For some patch-level prediction tasks, the patch-level label only applies to the center of the patch, while the area around the center provides useful context to the model for making a prediction (Supplemental Figure 2a). Since the class token embedding of the ViT model is essentially an aggregate of image token embeddings (embeddings corresponding to 16×16 pixel "tokens" within the original input patch, Supplemental Figure 2b), we explored whether the center token embeddings can be used to provide additional predictive value. Specifically, we explored whether an average pooling[6] of the center N×N token embeddings would improve the performance of our linear probe tasks, either concatenated with the standard class token embedding or used in isolation. Note that this variation was not part of the SSL training itself, but represents novel utilization of the output of ViT-based SSL models.

## SSL Model Evaluation

### Overview

Model evaluation was composed of four steps including both patch-level and weakly supervised slide-level or case-level tasks. First we established a benchmark set of 11 patch-level tasks to be used for both model tuning and linear evaluation. This included nine tissue-specific tasks (over six unique tissues) and two multi-tissue classification tasks, together spanning different optimal magnifications and task types (Table 1). Second, we evaluated models via linear probing on mitotic figure identification in melanoma and breast cancer, representing held-out patch-level tasks (not included in SSL model tuning). Third, to further evaluate generalization of the representations learned by the SSL models, we performed linear evaluation on a diverse set of additional held out, weakly supervised tasks. Fourth, we evaluated data efficiency and performance when fine-tuning end to end with SSL pre-training for two patch-level tasks.

## Patch-level linear probing

We selected a set of 11 patch-level tasks for linear probe evaluation (Table 1) with rigorously curated data sets spanning a diversity of task types, tissue types, and magnification requirements. Most of the datasets have been previously described[38–45] (with references also included in Table 1). The cervical dysplasia data is a pathologist annotated dataset from the same tertiary teaching hospital as the breast cancer tasks. Because TCGA is the primary training datasource for our SSL models, when necessary, new train/tune/test splits were generated in order to create test splits that did not include any TCGA data. The one exception is the TCGA study type task, for which the SSL train/tune/test splits were used (ie. no cases used for SSL training were included in the study type test set) .

These 11 tasks were used to compute a single aggregated patch-level “linear probe metric”. Linear probing is a commonly used approach for evaluating SSL models that involves fitting a linear model on SSL model patch embeddings to predict patch-level labels. All tasks were framed as classification tasks and evaluated using the area under the receiver operating characteristic curve (AUC) achieved by a regularized logistic regression model. For tasks with more than two classes, the macro-averaged AUC was used. The logistic regression models were initially trained on 10000 patches from the train split and evaluated on 5000 patches from the tune split. 5-fold cross-validation (stratified by slide) on the train patches was used to select L2 regularization weights. Since task performance is dependent on magnification, each task was evaluated at three magnification levels (5×, 10× 20×) and the highest AUC across magnifications was chosen. Finally a weighted average of these best task-specific AUCs was used to arrive at a composite “linear probe metric”. Task weights were selected to distribute weights over tissue types and tasks (Supplemental Table 2).

Linear probe metrics on the tune sets were used to guide model selection. Tune set metrics were used to select two final embedding approaches, using SimCLR and MSN, respectively. Then, individual training variations (see Methods for individual variations below) were evaluated and those achieving improvements of greater than three standard deviations over the respective baseline model were selected for combined optimization (Supplemental Table 3). Optimization of combined individual improvements led to selection of the “best” models (SimCLR-best and MSN-best).

Linear probe metrics on the test sets were calculated using 5000 patches with logistic regression model training on 10000 train set plus 5000 tune set patches (15000 patches total). Ablation experiments exploring the impact of various individual parameters were also performed via linear evaluation using the test sets.

## Mitotic figures and center embeddings

To evaluate generalization of these models to a "high magnification task" we used mitotic figure detection in breast cancer and melanoma, for which computer vision approaches often perform optimally using input patches at 40× magnification[38,46,47]. These tasks were not included in the main linear probe benchmark to avoid biasing towards optimization of these particular tasks which had lower performance and higher variability than the other tasks in initial exploratory experiments. The data used for mitotic figure detection in breast cancer has been described previously[38] with brief details in Supplemental Table 4. The dataset for melanoma represents an additional dataset from the same tertiary teaching hospital and with the same annotation method as for the breast cancer dataset, consisting of 175 cases split into train:tune:test sets in a 2:1:1 ratio. Due to the nature of the task (high magnification), the use of the center embeddings (see Center Embeddings section in Methods) was also utilized when evaluating mitotic figure detection.

Linear evaluation for mitotic figure detection was conducted similarly to the patch-level linear probe described above, with an average AUC over both tissue types calculated for each model.

## Weakly supervised linear evaluation

Models were further evaluated on weakly supervised tasks using available slide-level or case-level labels. These weakly supervised tasks include gene expression, survival prediction, TCGA study type, ER status for breast cancer, EGFR mutation status for lung cancer, and MSI status for colorectal cancer. TCGA images and data were used for all weakly supervised tasks reported here with the goal of reporting benchmark weakly supervised tasks that can be evaluated by other research groups as well. The same case-level splits for the SSL training were used here, such that the weakly supervised test sets represent held out cases not used for model development. Based on preliminary experiments, 1000 patches per case were sampled randomly and the corresponding patch-level embeddings from the SSL models were aggregated for use in linear evaluation. (Note that most TCGA cases have a single FFPE WSI, but there are cases with more than one FFPE WSI). Average pooling with linear evaluation was used in this work as an initial strategy to evaluate model representations independently from optimization of different possible slide-level or case-level models. All weakly supervised tasks were framed as classification tasks and were evaluated using AUC.

For gene expression prediction tasks, we focused on sets of genes for which feasibility of morphology based gene expression prediction in breast cancer [48,49] and liver cancer [50,51] has been reported previously. AUC was calculated per gene for binary classification, using per gene median expression as the threshold (RSEM-based RNAseq data as available via TCGA). The AUCs for each gene were averaged over defined gene sets as listed in Supplemental Table 5). Gene expression was only evaluated within the relevant TCGA study type (eg. breast cancer gene sets were only evaluated using TCGA BRCA data).

For ER status, AUC was calculated for classification of ER positive or negative using TCGA BRCA cases based on biomarker status labels as described previously[52]. For survival prediction tasks, AUC for predicting survival greater than 5 years was calculated and averaged across 8 TCGA study types with at least 50 examples in each class (KIRC, SARC, BRCA, LGG, HNSC, UCEC, LUSC, SKCM). For weakly supervised TCGA study type prediction, we evaluated macro-averaged AUC for the same 10 TCGA studies described in Table 1 for the patch-level version of the task and separately for all 32 TCGA studies. For EGFR status prediction, the AUC was calculated for presence of a pathogenic EGFR mutation in LUAD cases, using the somatic mutation calls available in TCGA[53] and manual pathologist annotation for pathogenicity. For MSI status, the AUC was calculated for classification of MSI-H using available MSI status from TCGA.

## Fine-tuning and data titration

To evaluate data efficiency of SSL pre-training and task-specific fine-tuning, we performed experiments for two well established tasks: prostate cancer Gleason grading in needle core biopsies and detection of metastatic breast cancer in lymph nodes [39,42,45]. Fine tuning was performed using either supervised ImageNet-21K pre-training or SSL pre-training (SimCLR-best and MSN-best) using different fractions of available WSIs for training (0.125, 0.25, 0.5, 1.0). The number of slides in the full train splits for these tasks is shown in Table 1 (Gleason NCB and CAMELYON16). For each titration point, five random subsamples of the training set WSIs were sampled. For each of the five subsamples per data titration point, a separate model was fully fine-tuned on 5 million sampled patches (class-balanced), resulting in 5 models per data titration point. Testing was performed using the 10× (~1 µM/pixel) patches for the Gleason grading task and the 20× (~0.5 µM/pixel) patches for CAMELYON16. The hyperparameters for fine-tuning are summarized in Supplemental Table 6.

## Metrics and Statistical analysis

Confidence intervals for linear probe and weakly supervised tasks were generated via bootstrap resampling over test set slides (as a more conservative approach than resampling over patches) with 1000 replicates. Linear probing used 5000 test set patches per task for each replicate. Weakly supervised evaluation used average pooling of embeddings for 1000 patches per slide or case with bootstrapping over test set slides. For tasks with more than two classes, AUCs for computing the linear probe metric were calculated using macro-averaging. Linear classifiers were trained using Scikit-Learn's LogisticRegression implementation, with L2 regularization (inverse coefficient selected via 5-fold cross validation from a grid of 10 log-spaced values ranging from 1e-4 to 1e4), using the L-BFGS optimizer trained for 100 updates. Analyses were performed using Python, Numpy, and scikit-learn libraries.

# Results

## Patch-level linear probing

Baseline models as well as SimCLR-best and MSN-best (combining individual optimizations) were evaluated on the test sets for all linear probe tasks. See Table 2 for linear probe metric performance for all models. For the optimized models, average AUC was 93.20% [95% CI 92.71%-93.72%] for SimCLR-best and 93.43% [95% CI 92.88%-93.84] for MSN-best. The SSL baseline models demonstrated average AUCs of 92.69% [95% CI 92.08%-93.18%] for SimCLR-base and 92.80% [95%CI 92.19%-93.27%] for MSN-base. Recently described approaches from Kang et al. (DINO ViT)[21] and Wang et al. (CTransPath)[23] were also implemented and demonstrated an average AUC of 92.00% [95% CI 91.34%-92.53%] (ViT-S/16) and 92.09% [95% CI 91.45%-92.62%] , respectively.

The best SSL models showed substantial increases in patch-level linear evaluation performance for all tasks over the supervised ImageNet baseline (Figure 2). Performance across all individual tasks, models, and magnifications is further summarized in Supplemental Figures 3 and 4. Generally the largest performance increases for SSL models were seen for tasks with relatively lower supervised ImageNet baseline performance, such as breast cancer nuclear pleomorphism, breast cancer tubule formation, lung adenocarcinoma histologic subtyping, and prostate cancer biopsy Gleason grading.

We also conducted ablation experiments in which we removed individual variations from the best performing models. For SimCLR-best, this meant removing either RandStainNA or the 512 patch size variant. For MSN-best, this meant removing RandStainNA, the SimCLR hybrid loss, or the Poisson noise component of PathBlur (Table 3). The one variation that consistently resulted in decreased test set performance when removed individually was RandStainNA (with default color perturbations still used), supporting the persistent value of this augmentation during training. Other variations improved performance in combination on the tune set, but the improvements did not clearly generalize to the test set. Further effort to understand generalization and task specific benefits of the different training variations is warranted.

## Mitotic figure identification and center embeddings

To evaluate use of SSL embeddings for a “held out” patch-level task (ie. independent from the linear probe tasks), we performed linear evaluation for mitotic figure detection in two cancer types: breast cancer and melanoma. This task requires high-magnification input patches and has the property that the patch-level labels only apply to a small center region of the input patch. This motivated us to evaluate the usefulness of “center embeddings”, which demonstrated substantial improvements over use of the “class token embeddings” alone (see “Center Embeddings” section of Methods). With the MSN-best model, the use of center-embeddings achieved an average AUC of 97.80% on the linear probe metric versus 89.06% without center embeddings (Table 4). For SimCLR-best, the improvement was even larger, from an average AUC of 79.32% without center embeddings to 95.21% with center embeddings.

Of note, the SimCLR-base model outperformed the SimCLR-best model at the mitotic figure identification task, perhaps consistent with the fact that this task was not used during model optimization and selection of "best" configurations. As such, we also evaluated the impact of individual training variations on this task specifically. Interestingly a few variations which did not meaningfully improve the overall linear probe metric did show a benefit for linear evaluation of mitotic figure identification. For MSN, PathBlur improved the performance over the MSN-baseline configuration substantially, increasing the mitotic figure average AUC by 3.41% (from 81.73% to 85.14%). For SimCLR, both PathBlur and hard negative mining modestly improved the mitotic figure average AUC over the SimCLR-baseline configuration. For PathBlur the increase was 0.63% (from 87.15% to 87.78%) and for hard negative mining the increase was 0.67% (from 87.15 to 87.82%). Taken together, these results highlight the need to further understand the impact of training variations for different tasks.

We also evaluated the use of center embeddings on the 11 linear probe metric tasks and found that center embeddings increased performance modestly overall, with the largest improvement for lung cancer histologic subtyping (as summarized in Supplemental Table 7 and Supplemental Figure 5). Lastly, to demonstrate that it was indeed the center embeddings that improved performance, we explored the use of increasing center NxN token embeddings (up to the complete set of 14×14 tokens per patch or using the 2×2 center embeddings alone; Supplemental Table 8).

## Weakly supervised linear evaluation

To better understand utilization and generalization of the patch-level embeddings for slide-level or case-level tasks, we evaluated performance for several types of such weakly supervised tasks including gene expression, survival prediction, ER status, EGFR mutation status, MSI status, and TCGA study type. Linear evaluation results for these tasks (using average pooling of patch embeddings across 1000 patches per slide/case) are summarized in Table 5 and Supplemental Table 9, demonstrating substantially better performance using pathology-specific SSL models as compared to ImageNet pre-training alone. These experiments offer initial insights into the value of the embedding models for several challenging, weakly supervised tasks, providing evidence for meaningful representations and capabilities of foundation model embeddings in regard to prognostic and predictive biomarkers. Exploring the use of patch-level embeddings from SSL models for slide-level and case-level tasks using more complex aggregation methods (e.g. cluster quantitation[41] or DeepMIL[54]) remain as important avenues to explore in future work.

## Fine-tuning data efficiency

One of the key value propositions for pathology-specific embedding models is the ability to reduce the necessary volume of data for training performant models across multiple applications. As such, we explored data efficiency of fine tuning a representative SSL model (MSN-best) for two well-established benchmark tasks: Gleason grading in prostate biopsies (Gleason NCB), and metastatic breast cancer detection in lymph nodes (CAMELYON16). We evaluated the impact of training data volume by simulating different amounts of slides available for training (using a fixed number of 5M labeled patches while titrating the number of slides). These results are summarized in Figure 3.

Fine-tuning pathology-specific SSL models (MSN-best and SimCLR-best) provided a substantial benefit over fine-tuning a model pre-trained on ImageNet across all data volumes, with the larger improvements at lower data volumes. For both tasks, fine-tuning the MSN-best model using roughly a quarter of available slides was comparable to fine-tuning a model pre-trained on ImageNet using all available slides. Additionally, the SSL pretraining outperformed ImageNet pretraining when fine-tuning on all available slides.

# Discussion

In this work we aimed to provide a rigorous exploration and evaluation of self-supervised learning approaches in pathology. The resulting models achieve performant results across a diverse set of benchmark tasks. This builds upon recent efforts that have compared popular methods[21], explored data and model scaling with standard approaches[19,20,22], or developed specific architecture variations[23,24].

One important component of this work was the use of a variety of tasks with rigorously established datasets and annotations for SSL model optimization as well as evaluation. For patch-level tasks, this included 17 different tissue types, 12 different cancer types, and several different types of tasks such as tumor identification, grading, subtyping, and classification. These tasks involved high-quality patch labels, established in most cases via annotation by multiple subspecialist pathologists and use of majority vote (rather than a single pathologist annotation). We also demonstrate generalizability and value of domain-specific embeddings on weakly supervised tasks, an important extension of prior efforts to understand the utility of SSL models for patch-level classification. The results clearly support the value of pathology-specific ViT-based SSL models across a range of tasks, particularly as compared to natural image pre-training alone. Furthermore, we found that the resulting embeddings can be used in a data-efficient manner to provide superior results using far less data than with ImageNet pre-training and supervised fine-tuning. As such, this work represents the establishment of an initial foundation model for H&E images in histopathology that we plan to explore further for downstream applications.

The weakly supervised results observed in this study not only demonstrate generalizable value of the embeddings to challenging tasks for which models weren't specifically optimized, but also suggest the features extracted by the pathology-specific SSL model could enable scalable biomarker discovery efforts. In other words, the ability of embeddings from a single embedding model to demonstrate predictive value for multiple biomarker tasks across cancer types, including gene expression, prognosis, and tumor grading, suggests that the models have learned biologically meaningful histomorphological features. Combined with explainability efforts[18,41] such embeddings and the evaluation of associated image regions (as shown in Supplemental Figure 6) could enable identification of novel biomarkers and biological insights.

The exploration of dataset size represents another notable aspect of this work, with preliminary experiments utilizing up to 600 million patches from the TCGA dataset. For context, other recent and rigorous benchmarking efforts report results using approximately 15 -100 million patches from TCGA[18,21,23]. Interestingly, although larger dataset size for training can be an important aspect of self-supervised learning[55], we observed that using 600 million versus 60 million patches did not have a significant impact on performance for our linear probe metric (thus, we used 60 million patches for final models and experiments). In addition, our experiments using larger models such as ViT-B only showed a small improvement, as seen in the test set metrics of Table 2. One possible explanation is "saturation" of performance for certain tasks in our linear probe metric (ie. some tasks may not be sufficiently "challenging" to reflect differences between models and some tasks have inherent inter-pathologist variability in the reference standard that confers performance ceilings for the metrics used). There may also be redundancy and overlap in features and patches during training such that added data doesn't substantially change the learned representations. Further work to understand the interaction between dataset size, optimal data sampling, model size, training variations, hyperparameters, and usefulness for specific downstream tasks remains a valuable area for further investigation.

While we explored a number of model variations (pathology specific color and blur augmentation, data balancing via clustering, reweighted losses, crop shape and size variations, patch size, model size, backbone architecture), we observed that only a limited number of these showed consistent benefits on the linear probe metric and combining individual improvements did not necessarily provide added benefit on external test set evaluation. One of the most consistently beneficial improvements observed was for RandStainNA color augmentation, which produces color augmentations during training that are more in line with the range of variability in histopathology. This result is consistent with similar benefits also reported recently[21], further suggesting it should be strongly considered when developing SSL models in histopathology. Other variations provided modest stand-alone improvements in exploratory experiments using the linear probe metric (eg. cluster-based data sampling, pathology specific blur, and patch size of 512x512.) However, despite efforts to optimize training and hyperparameter search when combining these variations, the final models were on par or only slightly better than with single variations alone. Similarly, model variations had different impacts on different tasks, with some tasks seemingly saturating in performance. Future work to explore optimization of combined variations, additional novel variations, and optimization of specific variants for specific use cases could be helpful.

Our exploration of center embeddings also offers interesting insights for downstream applications. A common approach for applying pathology ML models is patch classification where WSIs are broken down into hundreds of thousands of smaller patches, on the order of 32×32 pixels, that are each assigned a label from either annotations or model inference. For each of these labeled patches, a larger, surrounding input patch with useful contextual information represents the image that is actually fed into the ML model. Our linear probing tasks all correspond to this application, whereby labels correspond to the center patch within an input image patch. In our work we have shown that in high magnification tasks, the embeddings of center patch tokens can be particularly informative. As the label may not always correspond to the "center patch", an extension of this work could be further leveraging of information within individual patch tokens embeddings for more efficient and accurate patch classification.

This work also raises many future directions for continued exploration and improvement. Notably, the differences between top performing models are relatively modest overall, especially when using an average across tasks to enable multiple comparisons. As such, work comparing embeddings from different models for specific applications, and understanding their strengths and weaknesses, will continue to be a useful avenue of research. Another direction likely to be important in histopathology is models that utilize lower resolution/magnification input images, such that useful embeddings can be generated without requiring the potentially expensive inference for thousands of embeddings per whole slide image. Embeddings from large models and high resolution patches can provide high quality information, but may be expensive to generate and store. Additionally, while patch aggregation by average pooling was deliberately used in this work to evaluate the quality of the embedding model in isolation from variations in slide-level, non-linear models, further optimization of strategies to combine patch-level embeddings across a slide may demonstrate even more exciting promise for biomarker prediction and development. This could involve well-established attention-based multiple instance learning (ABMIL) approaches, aspects of hierarchical information as explored by Chen et al.[24], embedding-based clustering and quantification as input features[41], transformer based aggregation layers[56], or other unique ways of combining information from embeddings across patches and magnifications. Future work is also warranted to explore use of priors such as tumor masks from existing models and to consider clinically meaningful thresholds and metrics for specific biomarker tasks. Expanding on the number of tissue types evaluated in downstream applications will also be valuable for understanding the generalizability and capability of such models. We limited our tissue type classification tasks to well annotated tissue types in our datasets with a sufficient number of independent cases for evaluation. This included both cancer and non-cancer specimens, but additional tissue types and disease entities could also be added. Finally, the lack of improved performance with increased training patches in our work was an interesting albeit negative finding. Continuing to explore methods to improve embedding models by increasing data remains a challenging and exciting opportunity, particularly as the ecosystem for computational pathology becomes more mature and collaborative. Lastly, with recent breakthroughs in combining vision encoders and large language models, the ability to combine embedding models such as this with robust text and multimodal representations, raise exciting opportunities for what's possible in computational pathology.

This work has its limitations. While the diverse set of tissue types and tasks used for linear evaluation is an important component, it still represents a limited number of tasks and the single, linear probe summary metric cannot summarize the performance for all potential downstream applications. As noted, some of the tasks in our linear probe metric may also saturate in terms of the performance achieved with different embedding models. We also acknowledge that the models were selected and optimized based on tune set data for the same linear probe metric tasks used for most of the patch-level evaluation. While most patch-level tasks utilize external test set data sources and the mitotic figure and slide level tasks represent independent evaluation tasks not used during SSL model development, further exploration and validation will be necessary for specific downstream applications. Additionally, different SSL approaches and augmentations may be optimal for different tasks, as highlighted by the mitotic figure experiments. Future work is warranted to understand the nuances of using different pathology foundation models for different tasks. Lastly, we used AUC to facilitate averaging across many tasks, but different task-specific metrics could be used for task-specific benchmarking and optimization.

# Conclusion

The models developed in this work represent high-quality pathology foundation models that can potentially be applied to many downstream pathology applications, and with the goal of balancing performance and model size for feasible and scalable implementation. Ideally, such foundation models will enable a range of use cases including efficient development of models for diagnostic and predictive tasks, quality assurance and pre-analytical workflow improvements, tissue indexing and curation, and biomarker discovery and validation.

# Acknowledgments

We thank Boris Babenko and Michael Howell for critical feedback on the manuscript. We thank the Google Research team for software and hardware infrastructure support, especially

Mathilde Caron for her assistance with DINO and MSN as well as Rory Pilgrim and Kenneth Philbrick for helpful technical discussions. The results shown here are in part based upon data generated by the TCGA Research Network: https://www.cancer.gov/tcga.

# Tables

## Table 1: Linear probe tasks and datasets

| Task name | Description | Classes | Tissue type | Task type | Slides (train/tune/test) | Datasource reference |
|---|---|---|---|---|---|---|
| Breast IC | Breast invasive carcinoma detection | Benign, Invasive Carcinoma, DCIS | Breast | Tumor detection | 573/288/669 | Jaroensri et al. |
| Breast NP | Breast cancer nuclear pleomorphism grading | NP1, NP2, NP3 | Breast | Tumor grading | 681/343/945 | Jaroensri et al. |
| Breast TF | Breast cancer tubule formation grading | TF1, TF2, TF3 | Breast | Tumor grading | 681/343/945 | Jaroensri et al. |
| CAMELYON16 | Breast cancer detection in lymph nodes | Tumor, Non-Tumor | Lymph node | Tumor detection | 216/54/258 | Liu et al.<br>Benjordi et al. |
| Lung AD | Lung adenocarcinoma histologic subtypes | Acinar, Cribriform, Lepidic, Micropapillary, Papillary, Solid, Leukocyte, Necrosis, Non Tumor | Lung | Tumor subtyping | 73/25/50 | Sadwhani et al. |
| CIN | Cervical dysplasia grading | Non-tumor, CIN 1, CIN 2+ | Cervix | Tumor grading | 329/74/229 | N/A<br>(see methods) |
| CRC | Colorectal carcinoma detection | Tumor, Non-tumor | Colon and Rectum | Tumor detection | 149/51/44 | Wulczyn et al. |
| Gleason NCB | Prostate cancer Gleason grading on needle core biopsies | Benign, GP3, GP4, GP5 | Prostate | Tumor grading | 178/85/88 | Nagpal et al. |
| Gleason RP | Prostate cancer Gleason grading on radical prostatectomies | Benign, GP3, GP4, GP5 | Prostate | Tumor grading | 550/259/202 | Nagpal et al. |
| Tissue type | Tissue Type | 16 tissue classes: Appendix, Breast, Cervix, Colon and rectum, Fallopian Tube, Gallbladder, Liver, Lymph node, Ovary, Placenta, Prostate, Skin, Thyroid, Upper GI, Uterus, Vas deferens | 16 tissue classes | Tissue type classification | 17319/6488/6719 | Weng et al. |
| TCGA study type | TCGA study type | 10 TCGA studies: BLCA, BRCA, COAD, HNSC, KIRC, LIHC, LUAD, LUSC, OV, STAD | 10 TCGA studies | Tissue type classification | 2952/1466/1489 | N/A<br>(see methods) |

Abbreviations: GP: Gleason pattern; IC: Invasive Carcinoma; NP: nuclear pleomorphism; TF: Tubule formation; DCIS: Ductal carcinoma in situ; CIN: Cervical intraepithelial neoplasia. Note, the only test set containing TCGA data is for the TCGA study type task (see Methods).

**Table 2: Performance evaluation for patch-level linear probe tasks**

| Method | Architecture | Linear Probe Metric [95% CI] |
|---|---|---|
| Supervised (ImageNet-21k) | Vit-S/16 | 87.79 [87.06 - 88.49] |
| CTransPath | CNN + SwinTransformer | 92.09 [91.45 - 92.62] |
| $DINO_{p=16}$ (Lunit) | Vit-S/16 | 92.00 [91.34 - 92.53] |
| $DINO_{p=8}$ (Lunit) | Vit-S/8 | 92.26 [91.65 - 92.70] |
| SimCLR baseline | Vit-S/16 | 92.69 [92.08 - 93.18] |
| MSN baseline | Vit-S/16 | 92.80 [92.19 - 93.27] |
| SimCLR-best | Vit-S/16 | 93.20 [92.71 - 93.72] |
| MSN-best | Vit-S/16 | 93.43 [92.88 - 93.84] |
| SimCLR-best | Vit-B/16 | 93.08 [92.44 - 93.57] |
| MSN-best | Vit-B/16 | 93.74 [93.16 - 94.15] |

Average AUC across linear probe tasks for base and optimized models (test set metrics). Confidence intervals were generated via bootstrap resampling over test set slides with 1000 replicates.

**Table 3: Ablation of individual variants from SimCLR-best and MSN-best**

| SSL Method | RandStainNA | Patch Size (512×512) | Hybrid loss (0.5 weight) | PathBlur (Poisson only) | Linear Probe AUC |
|---|---|---|---|---|---|
| **SimCLR** | ✔ | ✔ | *NA* | *NA* | 93.20%* |
| | *X* | ✔ | *NA* | *NA* | 92.92% |
| | ✔ | *X* | *NA* | *NA* | 92.83% |
| **MSN** | ✔ | *NA* | ✔ | ✔ | 93.43%* |
| | ✔ | *NA* | ✔ | *X* | 93.40% |
| | ✔ | *NA* | *X* | ✔ | 93.51% |
| | *X* | *NA* | ✔ | ✔ | 93.29% |

Test set linear probe performance for optimized SimCLR-best and MSN-best models with ablation of individual components as indicated. For PathBlur, only Poisson noise was used in the MSN-best ablation experiments based on its superior performance on the linear probe metric for the tune set when compared to Poisson noise plus JPEG artifacts. *Top rows of each SSL method, represent the SimCLR-best and MSN-best models, respectively, as also reported in Table 2.

**Table 4: Performance for mitotic figure identification**

| Model | Average AUC for mitotic figure identification | |
|---|---|---|
| | **without center embeddings** | **with center embeddings** |
| Supervised (ImageNet-21k) | 73.77% | 81.74% |
| SimCLR-base | 87.82% | 96.64% |
| MSN-base | 82.63% | 97.27% |
| SimCLR-best | 79.32% | 95.21% |
| MSN-best | 89.06% | 97.80% |

Breast cancer and melanoma mitotic figure identification were evaluated separately using their respective test sets and the two resulting AUCs were averaged. Metrics represent evaluation at 20× (~0.5 microns/pixel), which was the best performing magnification for all mitotic figure models (of the three magnifications tested).

**Table 5: Weakly supervised task performance with linear evaluation**

| Method | Weakly-supervised Task (train/tune/test case count) | | | | | |
|---|---|---|---|---|---|---|
| | **TCGA Study Type (2314/1167/ 1165)** | **Survival TCGA pancancer (785/379/359)** | **BRCA set 1 (525/267/260)** | **ER status (BRCA) (489/256/238)** | **EGFR (LUAD) (230/120/119)** | **MSI Status (COAD) (198/101/104)** |
| Supervised (ImageNet-21k) | 98.8% | 63.8% | 73.8% | 86.6% | 63.9% | 86.7% |
| SimCLR-base | 99.6% | 64.6% | 77.3% | 89.7% | 67.1% | 93.0% |
| MSN-base | **99.7%** | 64.1% | 77.2% | 90.8% | 72.7% | 93.3% |
| SimCLR-best | 99.7% | 65.6% | 77.9% | 91.2% | 68.6% | 93.1% |
| MSN-best | 99.6% | **67.7%** | **78.4%** | **91.3%** | 69.4% | **94.3%** |
| CTransPath | 99.4% | 66.2% | 75.8% | 90.4% | **73.0%** | 91.7% |
| DINO ViT-S/16 | 99.4% | 67.1% | 76.8% | 90.8% | 70.6% | 90.9% |
| DINO ViT-S/8 | 99.6% | 67.7% | 77.4% | 90.9% | 70.0% | 89.7% |

Linear evaluation of weakly supervised tasks using average pooling of embeddings from 1000 patches per slide. Number of slides used for training and testing the linear model for each task are provided as indicated. Values represent AUC for slide-level classification as defined for each task in the methods. Bold indicates the best model for a given task. Abbreviations: BRCA: breast cancer; ER; Estrogen Receptor; EGFR: Epidermal growth factor receptor; MSI: Microsatellite instability; LUAD: Lung Adenocarcinoma. BRCA set 1 is composed of 25 genes based on the top image-based gene expression predictions from Wang et al., see Supplemental Table 6.

# Figures

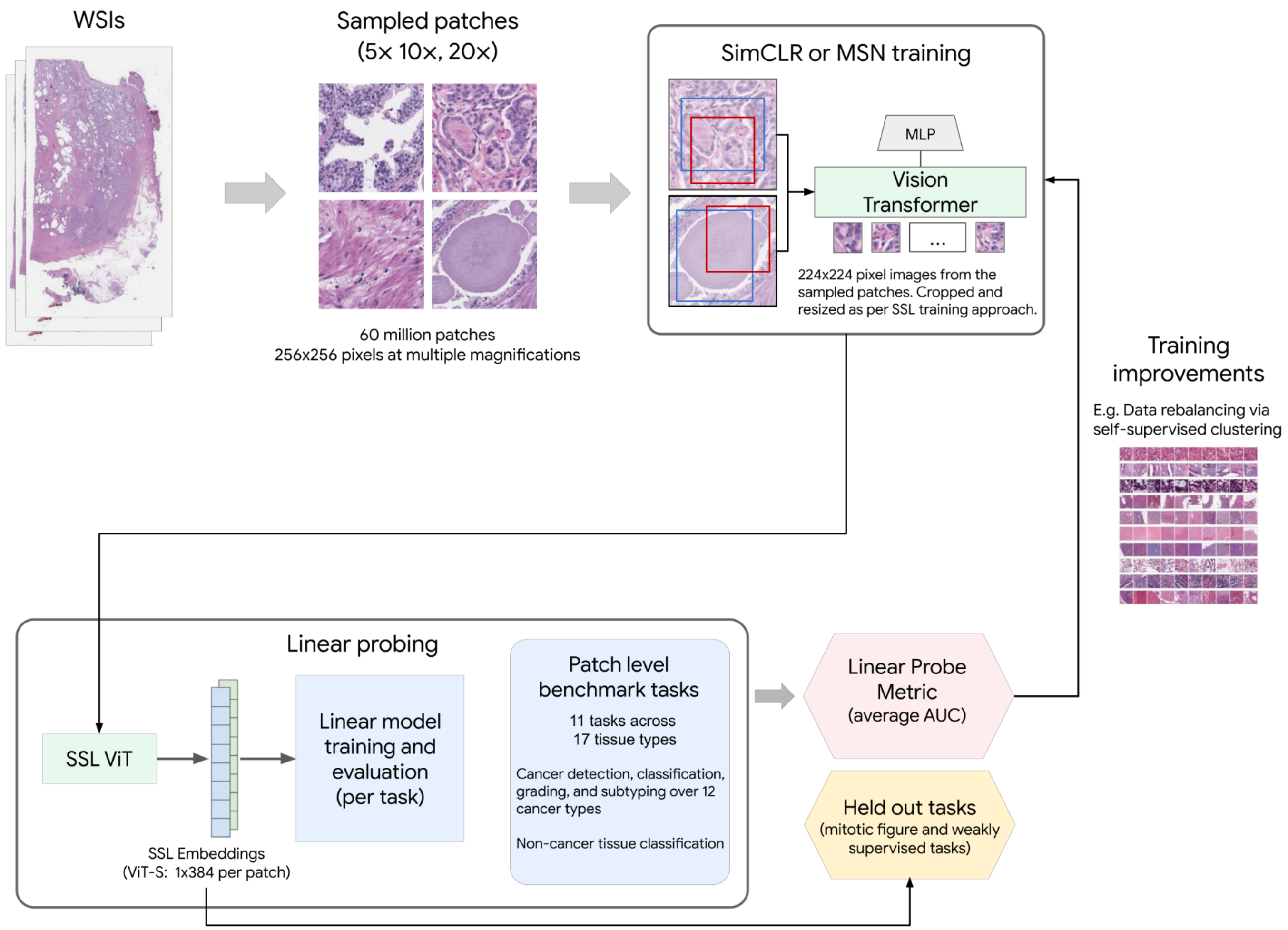

**Figure 1: Modeling and evaluation overview**

Schematic of modeling overview. Histopathology patches were sampled at multiple magnifications for use in exploration of training SSL models with different approaches. Resulting models were evaluated via linear probing on a diverse set of tasks and tissue types. The linear probe metric performance on a tune split was used to select and optimize training improvements to implement for final models. Final models were evaluated on test sets for the patch-level linear probe tasks as well as further evaluation on mitotic figure identification and weakly supervised classification and biomarker tasks.

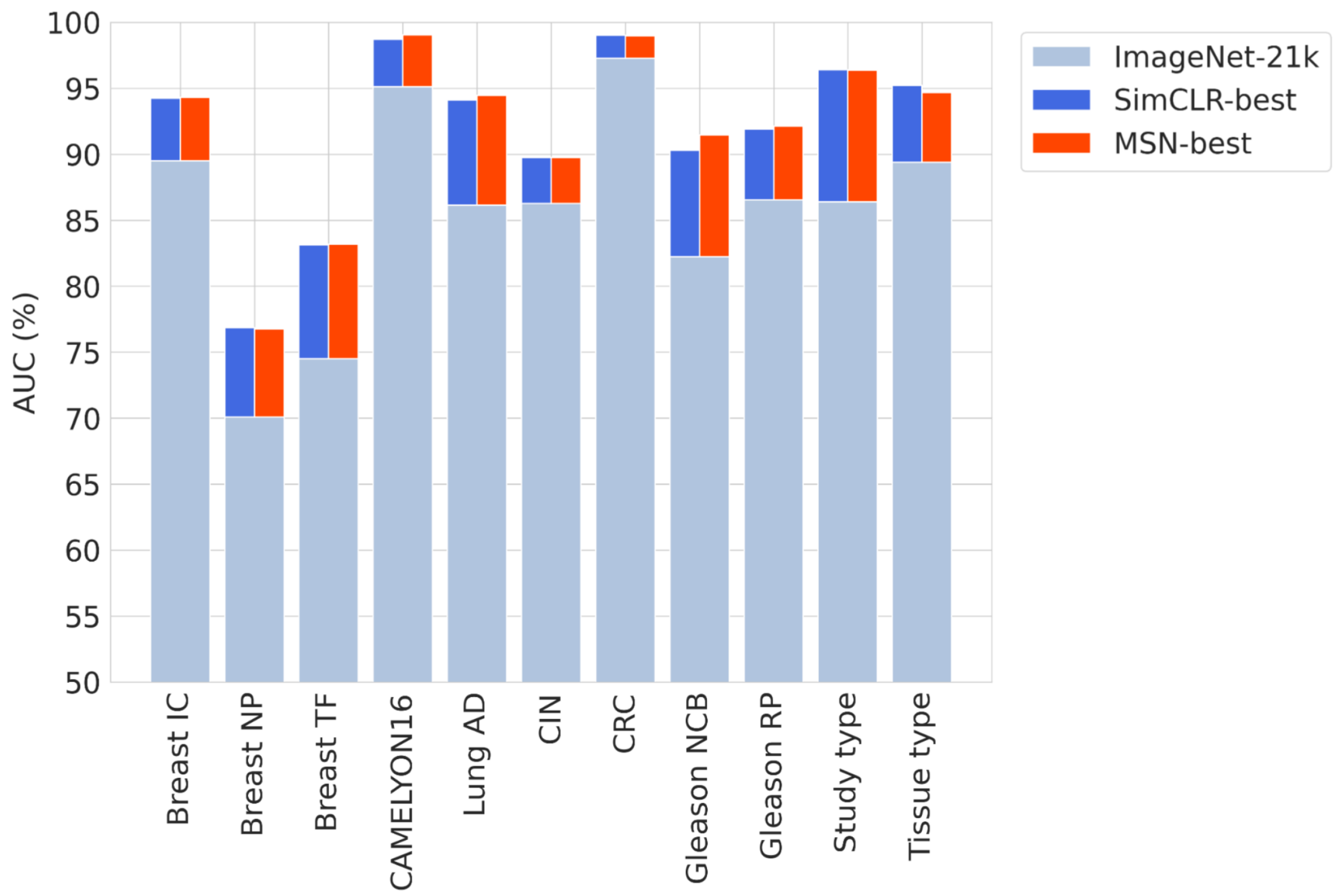

**Figure 2: Performance across patch-level linear probe tasks**
The AUC for each patch-level task of the linear probe metric is shown. The best performing magnification for each task for each model was used for comparison. Gray bars represent performance using embeddings from the supervised, ImageNet-trained model, blue bars represent performance using embeddings from SimCLR-best and red bars represent performance using embeddings from MSN-best. See Table 1 for additional task details.

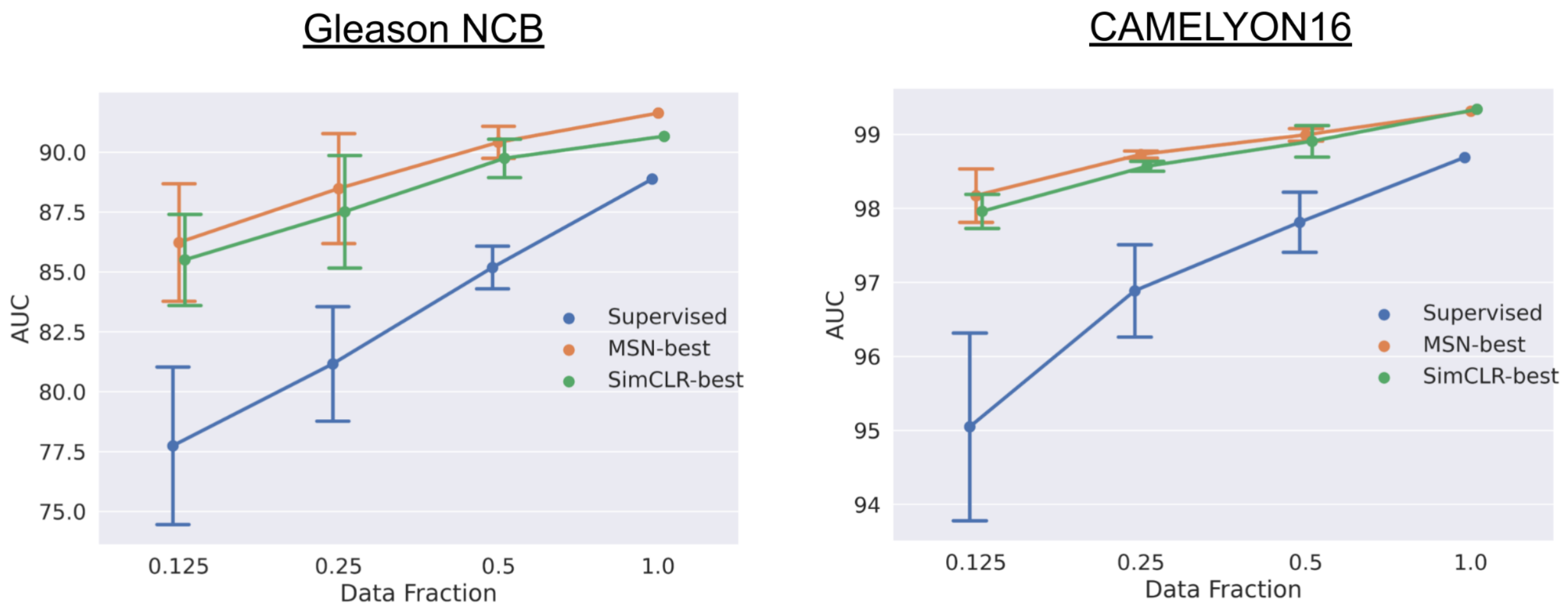

**Figure 3: Data efficiency with end-to-end fine tuning**
The AUCs for task-specific, fully fine-tuned models using different fractions of available whole slide images (WSIs) for training are plotted. Training for all WSI data fractions used the same number of labeled patches but a different number of WSIs (as described in the Methods). Error bars are standard deviations of five different models, each trained on a different subsample of the available WSIs for the corresponding data titration point. Supervised refers to the ImageNet-21k pretraining (ViT-S/16 architecture) and MSN-best and SimCLR-best represent the pathology domain self-supervised models.

# Supplemental Tables

## Supplemental Table 1a: Training details and hyperparameters for SSL base models

| **Shared hyperparameters** | **SimCLR/MSN-base** |
|---|---|
| Backbone | ViT-S/16 |
| Pretrained checkpoint | AugReg ImageNet-21k |
| Original image patch size | 256 |
| Color distortion (brightness, contrast, saturation, hue) | 0.4, 0.4, 0.4, 0.1 |
| Projection head | Three layer network with dimensions of 2048x256x1024 (input to output). |
| Batch size | 1024 |
| Optimizer | AdamW |
| Learning rate | 1e-4 (SimCLR) or 1e-3 (MSN) with linear warmup and cosine decay to 1e-6. |
| Weight decay rate | Initial rate of 0.04 with reverse cosine decay to 1. |
| Training epochs | 5 |
| | |
| **SimCLR hyperparameters** | **SimCLR-base** |
| NT-Xent temperature | 0.1 |
| Image crop size (Inception-style) | 224 |
| Crop area percentage | Uniform over [30%, 100%] |
| | |
| **MSN hyperparameters** | **MSN baseline** |
| Global crop size | 224 |
| Global crop area percentage | Uniform over [30%, 100%] |
| Number of global views | 1 teacher, 1 student |
| Local crop size | 96 |
| Local crop area percentage | Uniform over [5%, 30%] |
| Number of local views | 10 student |
| Mask percentage | 25% |
| Sinkhorn Norm | Yes |
| Centering | No |
| Cross-entropy teacher temperature | 0.0125 |
| Cross-entropy student temperature | 0.1 |
| ME-MAX temperature | 0.1 |
| Student to teacher EMA rate | Initially 0.996 with cosine decay. |

## Supplemental Table 1b: Training details and hyperparameters for SSL individual improvements

| Improvement | Hyperparameter | Hyperparameter range | Final best value (independent of other improvements) |
|---|---|---|---|
| **RandStainNA** | Color jitter strength (brightness, contrast, saturation, hue) | Weak (0.2, 0.2, 0.2, 0.05), moderate (0.4, 0.4, 0.4, 0.1), strong (0.8, 0.8, 0.8, 0.2) | Moderate |
| | Color jitter probability | 0, 0.8, 1 | 0.8 |
| | Staining probability | 0, 0.8, 1 | 0.8 |
| | | | |
| **PathBlur** | Add Poisson noise | False, True | True |
| | Add JPEG artifacts | False, True | True |
| | | | |
| **Data rebalancing** | Number of clusters | [500, 1k, 2k, 4k, ...., 20k] | 18k |
| | Number of K-Means training samples | 1M, 2M | 1M |
| | Number of sampled curated patches for SSL training | 6M, 12M, 24M | 6M |
| | | | |
| **Overlap cropping** | Min. overlap area (%) | 0.01, 0.05, 0.1, 0.2 | 0.2 |
| | Input patch size | 256, 512 | 256 |
| | | | |
| **Hard negative mining** | Reweight beta | 0.01, 0.02, 0.05, 0.1, ..., 2.0 | 0.1 |
| | | | |
| **Hybrid loss** | Global only (not using local views) | False, True | True |
| | SimCLR loss weight | 0.5, 1 | 1 |

### Supplemental Table 2: Weights per task for weighted average linear probe metric

| Task name | Relative Weight | Tissue type |
|---|---|---|
| Breast Inv Car | 0.33 | Breast |
| Breast NP | 0.33 | Breast |
| Breast TF | 0.33 | Breast |
| CAMELYON16 | 1 | Lymph node |
| Lung AD | 1 | Lung |
| CIN | 1 | Cervix |
| CRC | 1 | Colon |
| Gleason NCB | 0.5 | Prostate |
| Gleason RP | 0.5 | Prostate |
| Tissue type | 0.5 | Multi |
| TCGA study type | 0.5 | Multi |

**Supplemental Table 3: Evaluation of individual training variations for optimizing a pathology trained SSL model (tune set)**

| | Linear probe avg AUC | | | |
|---|---|---|---|---|
| **Method** | **SimCLR** | **Delta vs baseline** | **MSN** | **Delta vs baseline** |
| Baseline (average of 5 runs) | 93.36% | -- | 93.22% | -- |
| ViT-B | 93.58% | 0.22% | 93.52% | 0.30% |
| Data sampling via clusters | 93.50% | 0.14% | 93.30% | 0.08% |
| RandStainNA | 93.61% | 0.25% | 93.90% | 0.68% |
| PathBlur (Poisson + JPEG) | 93.46% | 0.10% | 93.56% | 0.34% |
| Overlap cropping | 93.68% | 0.32% | 93.42% | 0.20% |
| Add SimCLR loss | N/A | N/A | 93.55% | 0.33% |
| SimCLR loss w/ re-weighting | 93.41% | 0.05% | 93.66% | 0.44% |
| Patch size 512 | 93.68% | 0.32% | 93.25% | 0.03% |
| Center embeddings | 93.70% | 0.34% | 93.59% | 0.37% |

***Baseline model standard deviation: SimCLR-baseline: 0.035%; MSN-baseline: 0.061%. The Standard deviation for baseline models was calculated based on performance of 5 independent models trained using the respective baseline SSL approaches (with different random seeds). Improvement over baseline results are highlighted in gray if greater than 3 standard deviations above the respective baseline's average.

## Supplemental Table 4: Datasets for mitotic figure tasks

| Task name | Description | Classes | Tissue type | Slides (train/tune/test) |
|---|---|---|---|---|
| Breast MF | Breast mitosis detection | Negative, Mitotic Figure | Breast | 573/288/669 |
| Melanoma MF | Melanoma mitosis detection | Negative, Mitotic Figure | Skin | 310/93/107 |

Abbreviations: MF: Mitotic figure. Mitotic Figures were annotated and classified as described previously[38].

## Supplemental Table 5: Gene sets used for weakly supervised experiments

| **LIHC set 1 (Schmauch et al.)** | **LIHC set 2 (Xie et al.)** | **BRCA set 1 (Wang et al.)** | **BRCA set 2 (Mondol et al.)** |
|---|---|---|---|
| *CD3D* | *CYP3A4* | ADAM33 | BCL2 |
| *CD3E* | *CYP1A2* | *AURKA* | CCNE1 |
| *CD3G* | *GLUL* | *BIRC5* | CDC20 |
| *CD247* | *CYP2E1* | CCNB2 | CDCA7 |
| *CD19* | *FABP1* | CDC20 | CENPA |
| *MS4A1* | | CDC45 | CMC2 |
| *MKI67* | | CDCA5 | ESR1 |
| | | CDCA8 | FOXA1 |
| | | CENPA | KIF2C |
| | | DACT3 | MAPT |
| | | E2F2 | MLPH |
| | | KIF2C | MSANTD3 |
| | | KPNA2 | MYBL2 |
| | | MCM10 | NAT1 |
| | | MYBL2 | PGR |
| | | NCAPG | PTTG1 |
| | | NCAPH | SCUBE2 |
| | | NDC80 | SLC39A6 |
| | | ORC1 | SLC7A5 |
| | | PLK1 | UBE2C |
| | | PODN | |
| | | PRR11 | |
| | | SFRP2 | |
| | | SKA1 | |
| | | TROAP | |

**Supplemental Table 6: Hyperparameters for data efficiency experiments**

| Hyperparameter | Values |
|---|---|
| Magnification | CAMELYON16: 20×, Gleason NCB: 10× |
| Patch size | 224×224 |
| Batch size | 2048 |
| Image Perturbations/Augmentations | Rotate, Flip |
| Optimizer | Adam |
| Initial learning rate | 0.0001, 0.00001 |
| Learning rate decay steps | 100 |
| Learning rate decay rate | 0.9, 0.95 |
| Max train steps | 10000 |

**Supplemental Table 7: Test set evaluation of linear probe metric using center embeddings**

| Method | Architecture | Linear Probe Avg AUC (with center embeddings) | Linear Probe Avg AUC (without center embeddings) |
|---|---|---|---|
| ImageNet-21k | Vit-S/16 | 87.64% | 87.79% |
| SimCLR-base | Vit-S/16 | 93.12% | 92.69% |
| MSN-base | Vit-S/16 | 93.14% | 92.80% |
| SimCLR-best | Vit-S/16 | 93.51% | 93.20% |
| MSN-best | Vit-S/16 | 93.51% | 93.43% |

For center embedding linear evaluation, the average of the center 2×2 patch token embeddings was appended to the class embedding. Note that the last column reflects the same liner probe metric data for the corresponding models in Table 2.

**Supplemental Table 8: Linear probe metric with center embeddings usage**

| Center embeddings usage | Linear probe Avg AUC | Mitotic figure AUC |
|---|---|---|
| Baseline (no usage; CLS only) | 92.83% | 82.63% |
| 2×2 | 93.24% | 97.27% |
| 4×4 | 93.24% | 96.87% |
| 6×6 | 93.18% | 93.10% |
| 8×8 | 93.08% | 89.60% |
| 10×10 | 93.00% | 86.73% |
| 12×12 | 92.95% | 84.73% |
| 14×14 (all patch tokens) | 92.91% | 85.07% |
| 2×2 only (no CLS embedding) | 90.70% | 97.09% |

Results for variations in size of the region used for center embeddings, as evaluated on the tune split data. The CLS token embedding was concatenated to the center patch token embeddings, except for the last row, as indicated.

**Supplemental Table 9: Extended weakly supervised evaluation**

| Method | **BRCA set 2 (525/267/260)** | **BRCA-oncotype (525/267/260)** | **BRCA-PAM50 (525/267/260)** | **LIHC set 1 (180/88/90)** | **LIHC set 2 (180/88/90)** | **MSI Status (MSI-H and L combined) (COAD) (198/101/104)** | **TCGA Study Type (31 studies) (4642/2326/2329)** |
|---|---|---|---|---|---|---|---|
| Supervised (ImageNet-21k) | 73.5% | 67.1% | 69.7% | 67.3% | 65.5% | 77.7% | 98.5% |
| SimCLR-base | 77.1% | 69.7% | 73.3% | 71.4% | 67.4% | 78.4% | 99.5% |
| MSN-base | 76.8% | 70.2% | 73.0% | 68.9% | **69.6%** | 79.9% | **99.5%** |
| SimCLR-best | 78.1% | 70.0% | 74.3% | 72.1% | 67.7% | 79.5% | 99.3% |
| MSN-best | **78.2%** | **70.7%** | **74.6%** | 72.6% | 67.6% | 81.4% | **99.5%** |
| CTransPath | 76.0% | 69.4% | 72.6% | 71.3% | 66.8% | **81.7%** | 99.0% |
| DINO ViT-S/16 | 76.5% | 69.3% | 73.0% | **73.9%** | 67.2% | 79.4% | 99.2% |
| DINO ViT-S/8 | 76.8% | 70.1% | 73.4% | 73.0% | 67.3% | 79.4% | 99.4% |

As described for weakly supervised tasks, results represent linear evaluation using average pooling of embeddings from 1000 patches per slide. For gene expression tasks, values are AUC for case-level classification per gene, averaged across the genes in each gene set, as defined in the methods. For MSI, these data represent classification of MSI-H or MSI-L versus MSS (as opposed to MSI-H versus other as calculated for Table 5 of the main text). For study type, one study of the 32 TCGA solid tumor studies was dropped due to too few cases (DBLC). Bold indicates the highest AUC for a given task. Abbreviations: BRCA: breast cancer; LIHC; Liver hepatocellular cancer; MSI: microsatellite instability; COAD: colorectal adenocarcinoma. Oncotype and PAM-50 are well established gene sets in breast cancer classification and prognosis. BRCA-set 2, LIHC-set 1, and LIHC-set 2 gene sets are based on top predicted genes from prior work, see Supplemental Table 4 for genes and references.

# Supplemental Figures

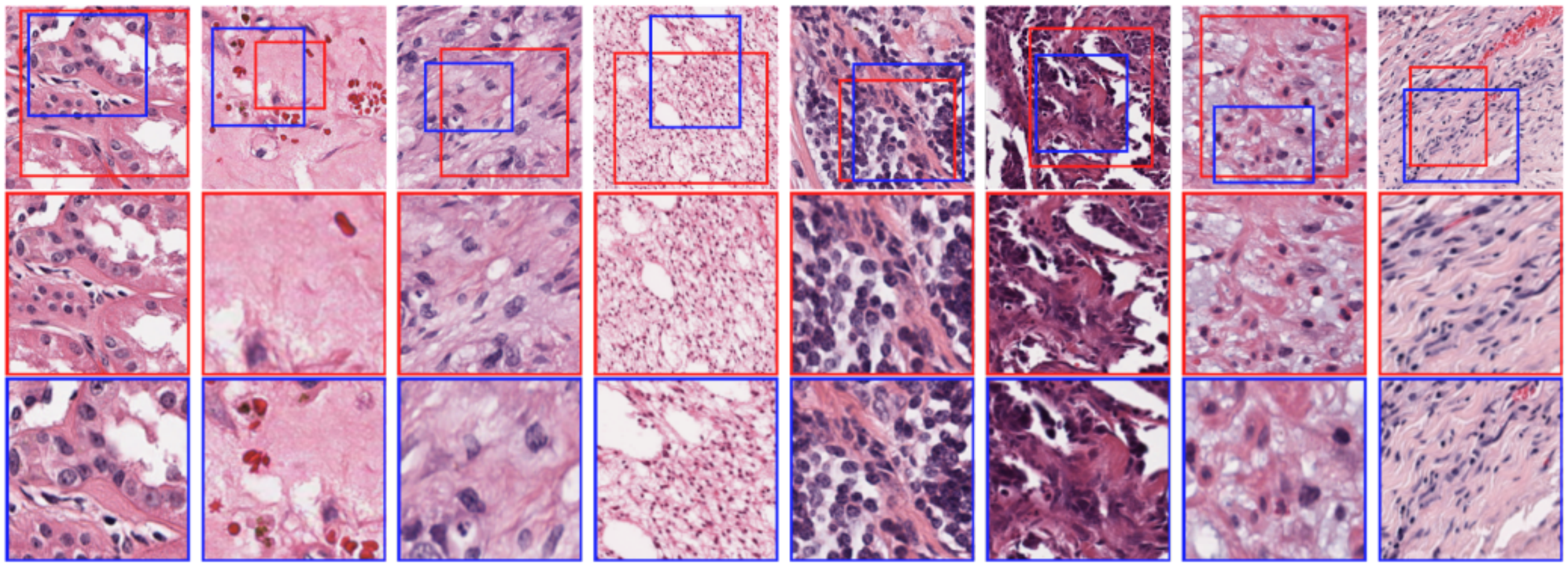

**Supplemental Figure 1: Visualization of overlapping input image crops**
Visualization of optimized input image cropping. [*top*] From within a sampled input patch, a global crop is randomly selected (red rectangle), followed by a local crop (blue rectangle), such that at least 20% of the area of the global crop is included within the local crop. The extent of overlap for optimal performance on the linear probe metric was found via grid-search. The global crops [*middle*] or local crops [*bottom]* are shown when resized to 224×224, to be used as input images for training the SSL model (before applying other data transformations).

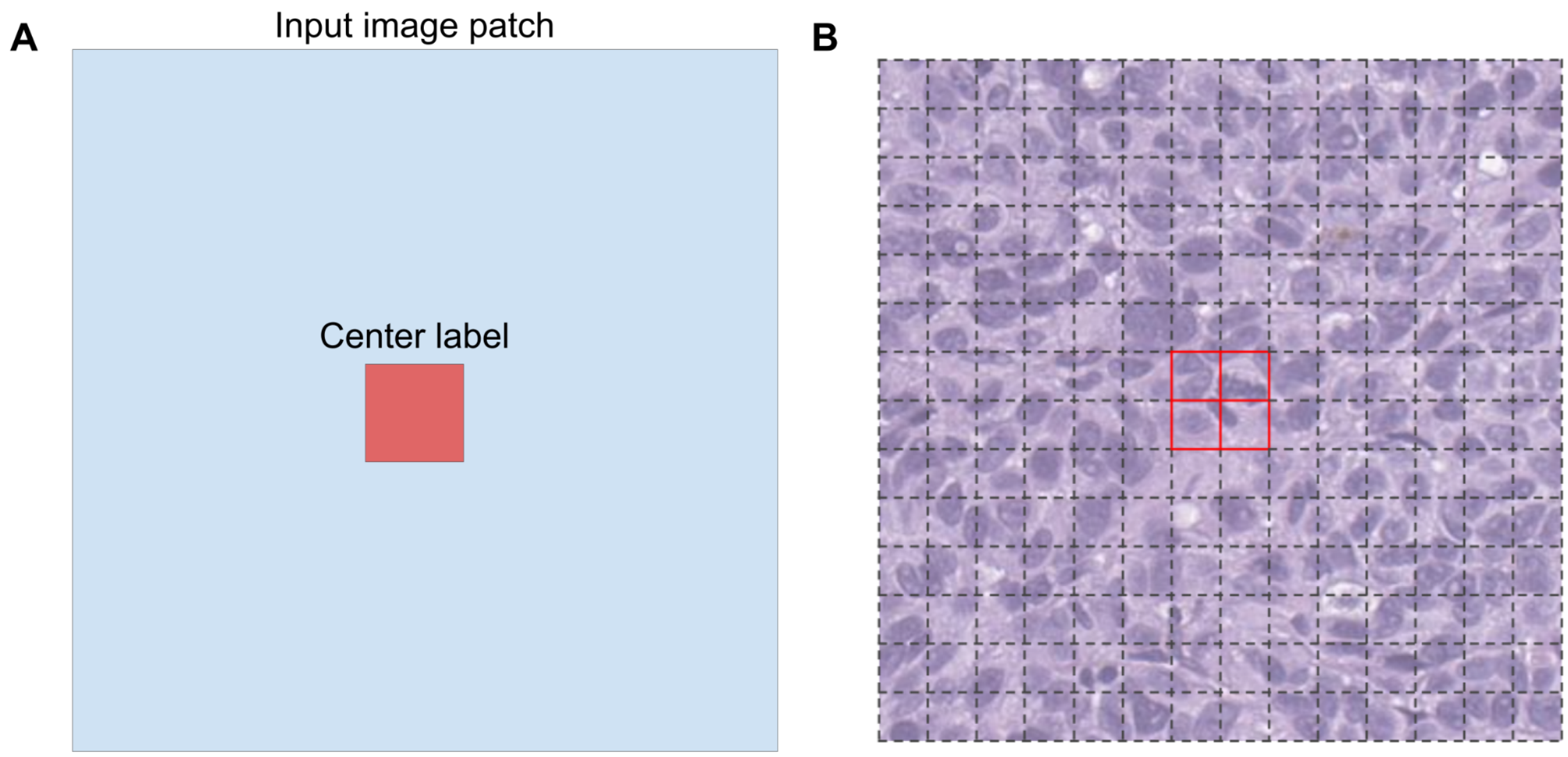

**Supplemental Figure 2: Center label and center embeddings visualization**

(A) The notion of an input image patch with a label that specifically applies to a small central region is depicted. In (B), an example 224×224 pixel input image patch is depicted with the dashed lines representing the input patch tokens to the ViT (16×16 pixels) that comprise each input image. Red outlines indicate the center 2×2 patch tokens. The output of the ViT contains embeddings for each of the input tokens as well as an embedding of a “class token” that represents the entire input patch

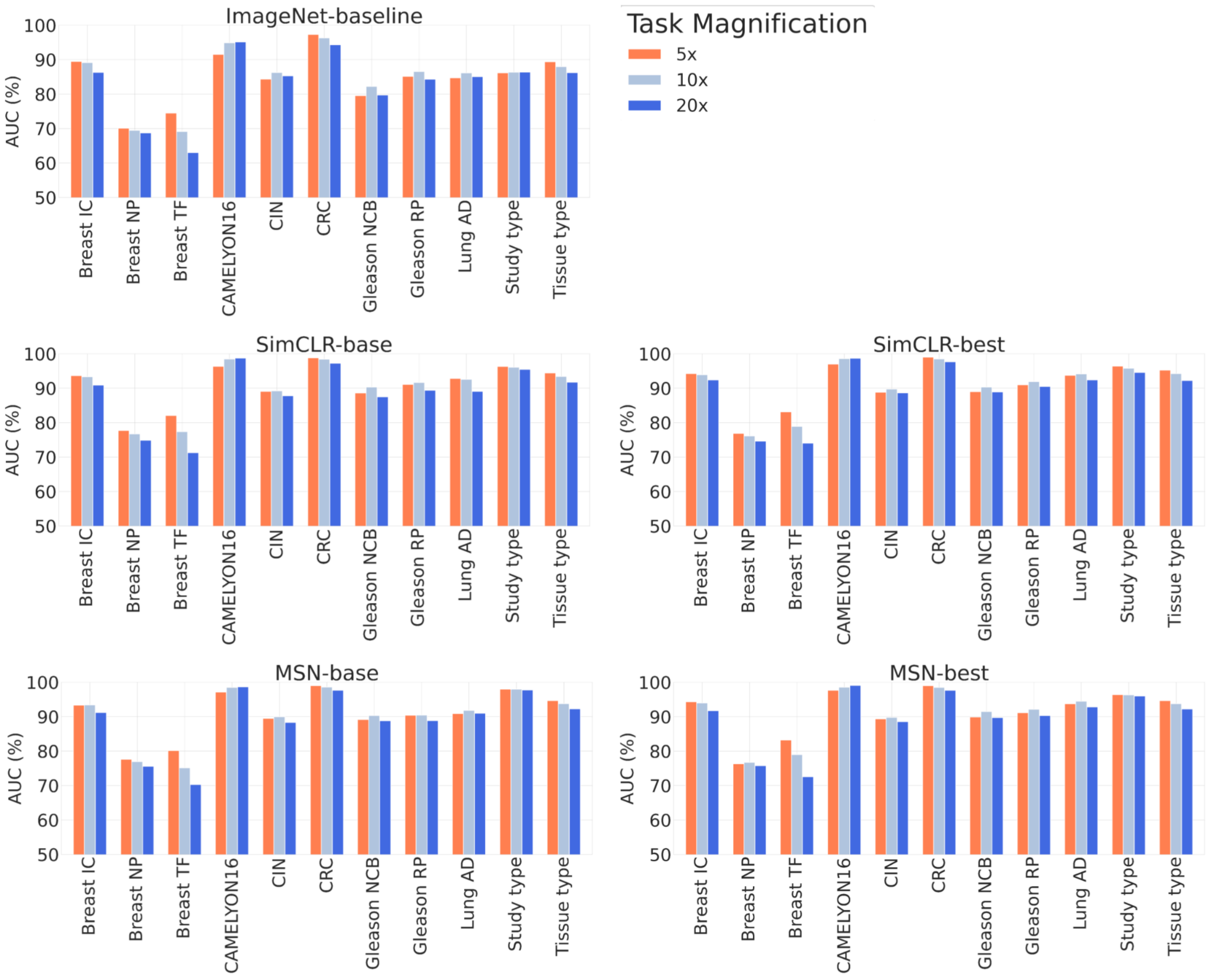

**Supplemental Figure 3: Performance across tasks and magnifications for multiple embedding models**

The AUC for each linear probe metric task is shown. Each panel represents a different model as indicated (ImageNet-21k pretrained, SimCLR-base, MSN-base, SimCLR-best, MSN-best). The different color bars represent the results when running inference and performing evaluation using patches at 5× (~2 μM/pixel), 10× (~1 μM/pixel), or 20× (~0.5 μM/pixel), respectively.

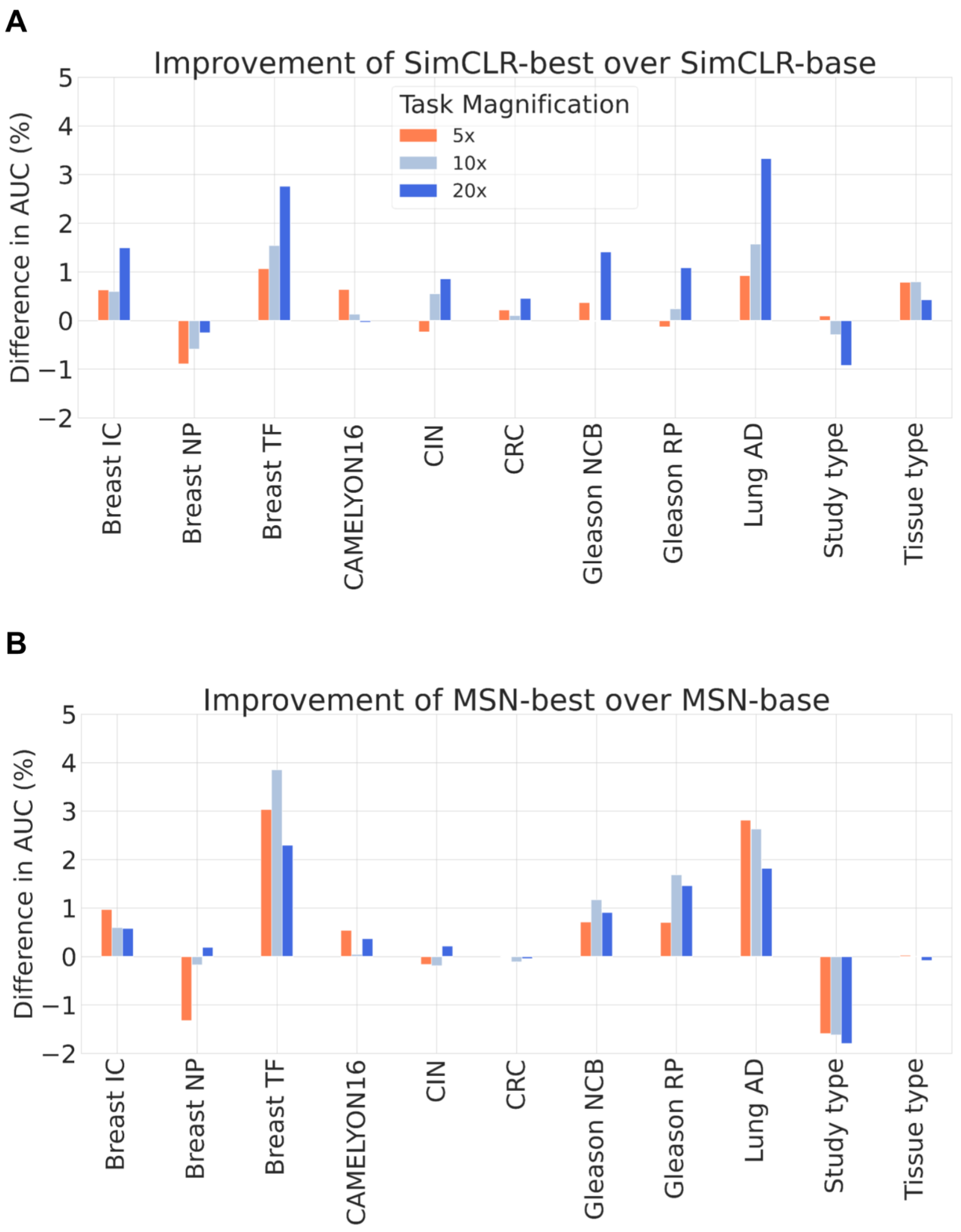

**Supplemental Figure 4: Performance difference between the baseline and selected SSL models across tasks**

The delta for the linear probe metric performance across tasks and across inference magnification is shown for (A) SimCLR-best versus SimCLR-base; and (B) MSN-best versus MSN-base. Related to Supplemental Figure 2 which shows the AUC values by task and magnification for the same models. The different color bars represent the results when running inference and performing evaluation using patches at 5× (~2 µM/pixel), 10× (~1 µM/pixel), or 20× (~0.5 µM/pixel).

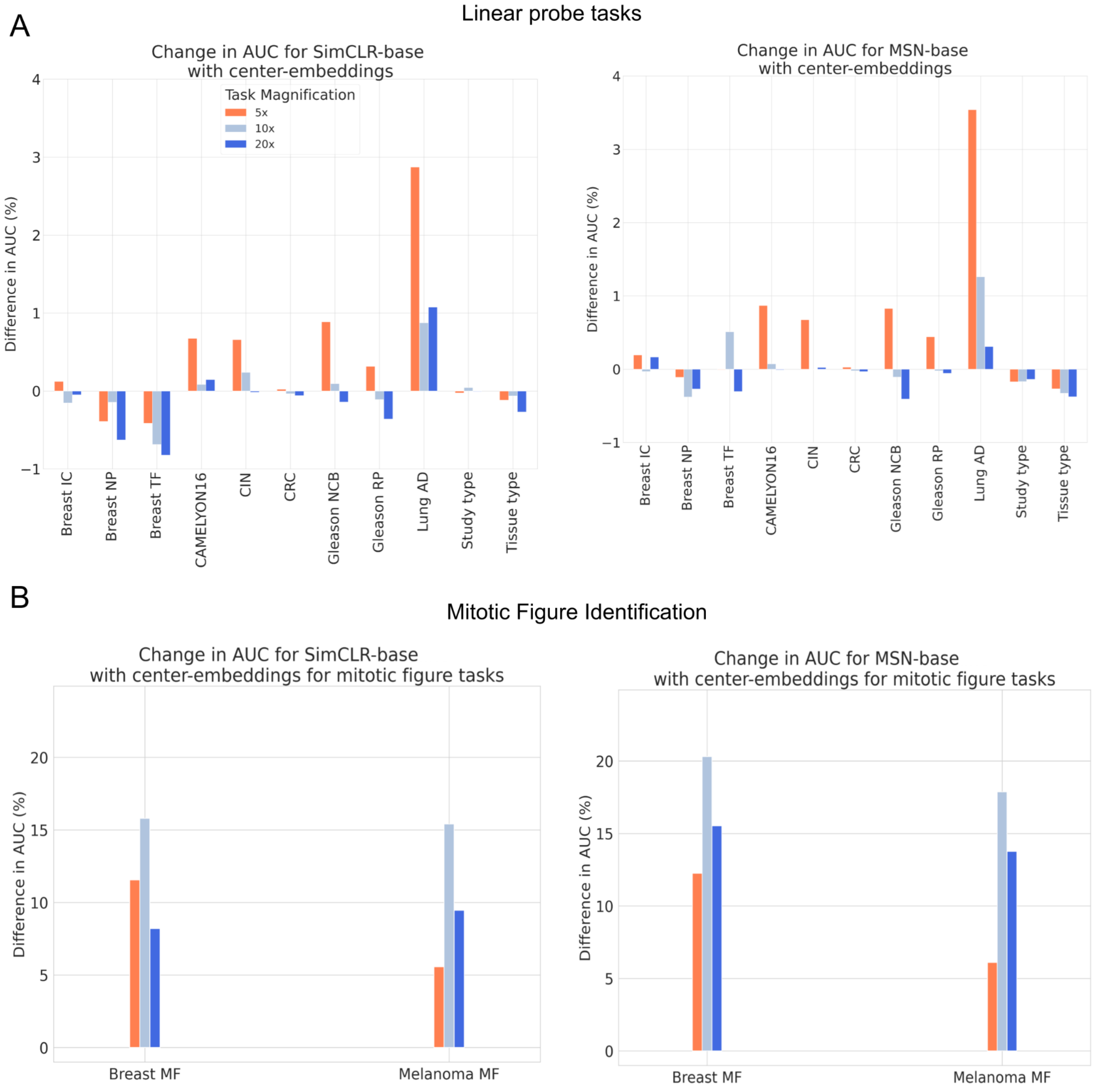

**Supplemental Figure 5: Performance of baseline models across tasks, with and without use of center embeddings**

Bar charts represent the delta in AUC across linear probe tasks (A) or mitotic figure tasks (B) when using the average of the center 2x2 embeddings appended to the class embedding versus using the class embedding alone. Note that while 10x shows the most pronounced improvements in the mitotic figure tasks with use of center-embeddings, 20x still demonstrated the highest AUC for this task, both with center embeddings and without. The different color bars represent the results when running inference and performing evaluation using patches generated at ~2 µM/pixel (5x), ~1 µM/pixel (10x), or ~0.5 µM/pixel (20x).

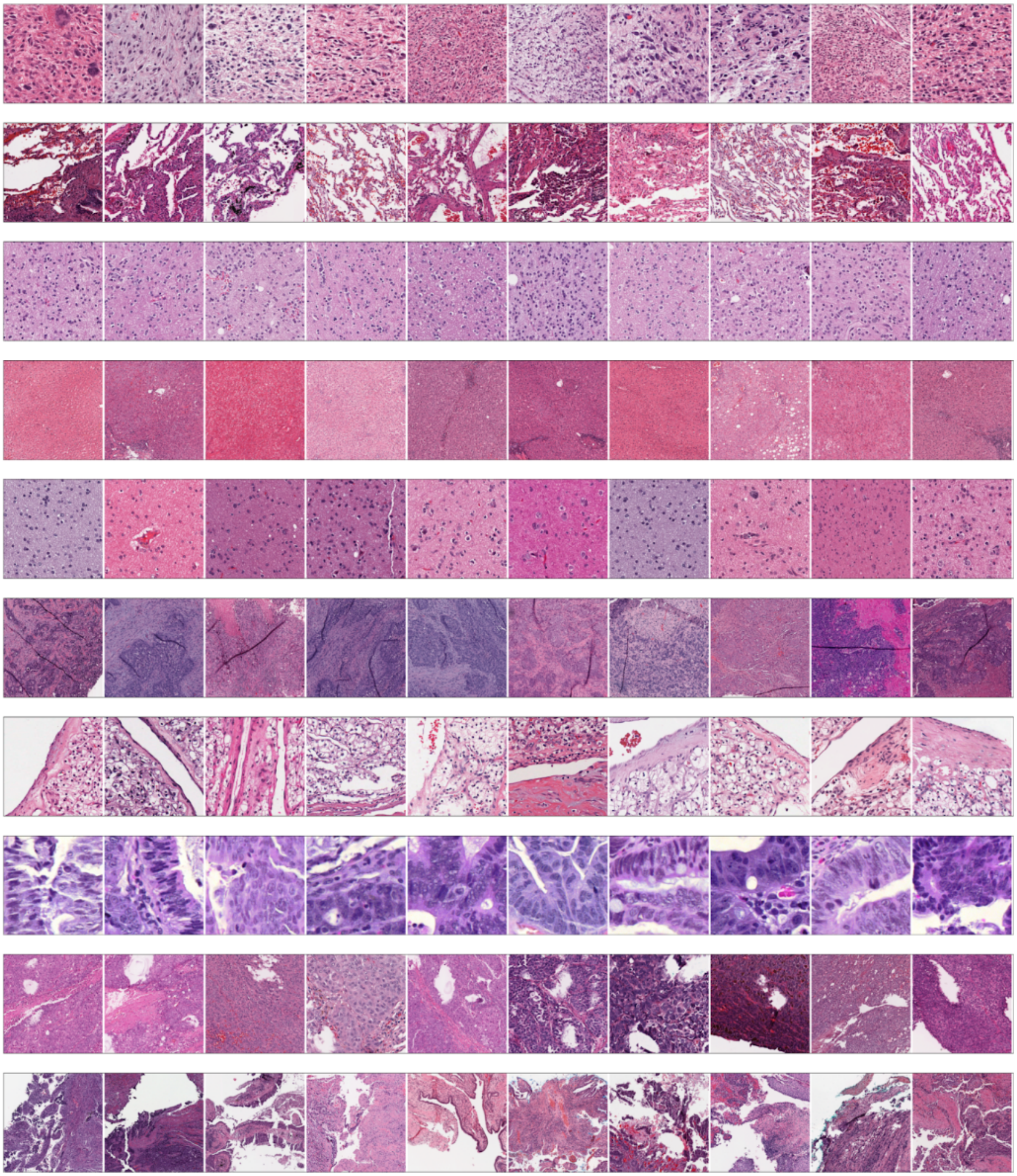

**Supplemental Figure 6: Example patches and clusters based on SSL embeddings**
Patches (n = 60 million) were clustered by k-means clustering (k = 18,000) based on their embedding vectors (using the SimCLR-best model in this example). Random patches from a random subset of clusters (n = 10) are shown here (one cluster per row) to provide visual examples. These examples illustrate that embedding vectors represent visually distinct features with high potential for downstream applications such as classification, feature evaluation, and similarity-based retrieval.

# References

1. Bera, K., Schalper, K. A., Rimm, D. L., Velcheti, V. & Madabhushi, A. Artificial intelligence in digital pathology - new tools for diagnosis and precision oncology. *Nat. Rev. Clin. Oncol.* **16**, (2019).
2. Echle, A. *et al.* Deep learning in cancer pathology: a new generation of clinical biomarkers. *Br. J. Cancer* **124**, (2021).
3. Steiner, D. F., Chen, P. C. & Mermel, C. H. Closing the translation gap: AI applications in digital pathology. *Biochim. Biophys. Acta Rev. Cancer* **1875**, (2021).
4. Cooper, M., Ji, Z. & Krishnan, R. G. Machine learning in computational histopathology: Challenges and opportunities. *Genes Chromosomes Cancer* **62**, 540–556 (2023).
5. Song, A. H. *et al.* Artificial intelligence for digital and computational pathology. *Nature Reviews Bioengineering* 1–20 (2023).
6. Caron, M. *et al.* Emerging Properties in Self-Supervised Vision Transformers. (2021).
7. Dosovitskiy, A. *et al.* An Image is Worth 16x16 Words: Transformers for Image Recognition at Scale. (2020).
8. Grill, J.-B. *et al.* Bootstrap your own latent: A new approach to self-supervised Learning. (2020).
9. He, K., Fan, H., Wu, Y., Xie, S. & Girshick, R. Momentum Contrast for Unsupervised Visual Representation Learning. (2019).
10. Assran, M. *et al.* Masked Siamese Networks for Label-Efficient Learning. (2022).
11. Chen, T., Kornblith, S., Norouzi, M. & Hinton, G. A Simple Framework for Contrastive Learning of Visual Representations. (2020).
12. van den Oord, A., Li, Y. & Vinyals, O. Representation Learning with Contrastive Predictive

Coding. (2018).

13. Huang, S.-C. *et al.* Self-supervised learning for medical image classification: a systematic review and implementation guidelines. *npj Digital Medicine* **6**, 1–16 (2023).
14. Azizi, S. *et al.* Robust and data-efficient generalization of self-supervised machine learning for diagnostic imaging. *Nature Biomedical Engineering* **7**, 756–779 (2023).
15. Fashi, P. A., Hemati, S., Babaie, M., Gonzalez, R. & Tizhoosh, H. R. A self-supervised contrastive learning approach for whole slide image representation in digital pathology. *J. Pathol. Inform.* **13**, (2022).
16. Ciga, O., Tl, X. & Martel, A. L. Self supervised contrastive learning for digital histopathology. *Machine Learning with Applications* **7**, 100198 (2022).
17. Koohbanani, N. A., Unnikrishnan, B., Khurram, S. A., Krishnaswamy, P. & Rajpoot, N. Self-Path: Self-Supervision for Classification of Pathology Images With Limited Annotations. *IEEE Transactions on Medical Imaging* **40**, (2021).
18. Chen, R. J. & Krishnan, R. G. Self-Supervised Vision Transformers Learn Visual Concepts in Histopathology. (2022).
19. Chen, R. J. *et al.* A General-Purpose Self-Supervised Model for Computational Pathology. *ArXiv* (2023).
20. Vorontsov, E. *et al.* Virchow: A Million-Slide Digital Pathology Foundation Model. (2023).
21. Kang, M., Song, H., Park, S., Yoo, D. & Pereira, S. Benchmarking Self-Supervised Learning on Diverse Pathology Datasets. in *Proceedings of the IEEE/CVF Conference on Computer Vision and Pattern Recognition* 3344–3354 (2023).
22. Filiot, A. *et al.* Scaling Self-Supervised Learning for Histopathology with Masked Image Modeling. *medRxiv* 2023.07.21.23292757 (2023) doi:10.1101/2023.07.21.23292757.
23. Wang, X. *et al.* Transformer-based unsupervised contrastive learning for histopathological

image classification. *Med. Image Anal.* **81**, 102559 (2022).

24. Chen, R. J. *et al.* Scaling Vision Transformers to Gigapixel Images via Hierarchical Self-Supervised Learning. (2022).
25. Steiner, A. *et al.* How to train your ViT? Data, Augmentation, and Regularization in Vision Transformers. (2021).
26. Chen, T., Kornblith, S., Swersky, K., Norouzi, M. & Hinton, G. Big Self-Supervised Models are Strong Semi-Supervised Learners. (2020).
27. He, K. *et al.* Masked Autoencoders Are Scalable Vision Learners. (2021).
28. Shen, Y., Luo, Y., Shen, D. & Ke, J. RandStainNA: Learning Stain-Agnostic Features from Histology Slides by Bridging Stain Augmentation and Normalization. (2022) doi:10.1007/978-3-031-16434-7_21.
29. Ruifrok, A. C. & Johnston, D. A. Quantification of histochemical staining by color deconvolution. *Anal. Quant. Cytol. Histol.* **23**, (2001).
30. Reinhard, E., Adhikhmin, M., Gooch, B. & Shirley, P. Color transfer between images. https://ieeexplore.ieee.org/document/946629.
31. Whole-Slide Image Focus Quality: Automatic Assessment and Impact on AI Cancer Detection. *J. Pathol. Inform.* **10**, 39 (2019).
32. Oquab, M. *et al.* DINOv2: Learning Robust Visual Features without Supervision. (2023).
33. Chen, T., Luo, C. & Li, L. Intriguing Properties of Contrastive Losses. (2020).
34. Robinson, J., Chuang, C.-Y., Sra, S. & Jegelka, S. Contrastive Learning with Hard Negative Samples. (2020).
35. Chuang, C.-Y., Robinson, J., Yen-Chen, L., Torralba, A. & Jegelka, S. Debiased Contrastive Learning. (2020).
36. Chen, X. & He, K. Exploring Simple Siamese Representation Learning. (2020).

37. Garrido, Q., Chen, Y., Bardes, A., Najman, L. & LeCun, Y. On the duality between contrastive and non-contrastive self-supervised learning. in *The Eleventh International Conference on Learning Representations* (2022).

38. Jaroensri, R. *et al.* Deep learning models for histologic grading of breast cancer and association with disease prognosis. *npj Breast Cancer* **8**, 1–12 (2022).

39. Liu, Y. *et al.* Artificial Intelligence-Based Breast Cancer Nodal Metastasis Detection: Insights Into the Black Box for Pathologists. *Arch. Pathol. Lab. Med.* **143**, (2019).

40. Sadhwani, A. *et al.* Comparative analysis of machine learning approaches to classify tumor mutation burden in lung adenocarcinoma using histopathology images. *Sci. Rep.* **11**, 1–11 (2021).

41. Wulczyn, E. *et al.* Interpretable survival prediction for colorectal cancer using deep learning. *NPJ Digital Medicine* **4**, (2021).

42. Nagpal, K. *et al.* Development and Validation of a Deep Learning Algorithm for Gleason Grading of Prostate Cancer From Biopsy Specimens. *JAMA Oncol* **6**, 1372–1380 (2020).

43. Weng, W.-H., Cai, Y., Lin, A., Tan, F. & Chen, P.-H. C. Multimodal Multitask Representation Learning for Pathology Biobank Metadata Prediction. (2019).

44. Nagpal, K. *et al.* Development and validation of a deep learning algorithm for improving Gleason scoring of prostate cancer. *npj Digital Medicine* **2**, 1–10 (2019).

45. Bejnordi, B. E. *et al.* Diagnostic Assessment of Deep Learning Algorithms for Detection of Lymph Node Metastases in Women With Breast Cancer. *JAMA* **318**, 2199–2210 (2017).

46. Veta, M. *et al.* Predicting breast tumor proliferation from whole-slide images: The TUPAC16 challenge. *Med. Image Anal.* **54**, (2019).

47. Aubreville, M. *et al.* Mitosis domain generalization in histopathology images - The MIDOG challenge. *Med. Image Anal.* **84**, (2023).

48. Wang, Y. *et al.* Predicting Molecular Phenotypes from Histopathology Images: A Transcriptome-Wide Expression–Morphology Analysis in Breast Cancer. *Cancer Res.* **81**, 5115 (2021).

49. Mondol, R. K. *et al.* hist2RNA: An Efficient Deep Learning Architecture to Predict Gene Expression from Breast Cancer Histopathology Images. *Cancers* **15**, (2023).

50. Schmauch, B. *et al.* A deep learning model to predict RNA-Seq expression of tumours from whole slide images. *Nat. Commun.* **11**, 1–15 (2020).

51. Xie, R., Pang, K., Bader, G. D. & Wang, B. Spatially Resolved Gene Expression Prediction from H&E Histology Images via Bi-modal Contrastive Learning. (2023).

52. Gamble, P. *et al.* Determining breast cancer biomarker status and associated morphological features using deep learning. *Communications medicine* **1**, (2021).

53. Ellrott, K. *et al.* Scalable Open Science Approach for Mutation Calling of Tumor Exomes Using Multiple Genomic Pipelines. *Cell systems* **6**, 271 (2018).

54. Ilse, M., Tomczak, J. M. & Welling, M. Attention-based Deep Multiple Instance Learning. (2018).

55. Goyal, P. *et al.* Self-supervised Pretraining of Visual Features in the Wild. (2021).

56. Wagner, S. J. *et al.* Transformer-based biomarker prediction from colorectal cancer histology: A large-scale multicentric study. *Cancer Cell* **41**, 1650 (2023).